# Recovering the spectral and spatial information of an object behind a scattering media

LEI ZHU, JIETAO LIU, LEI FENG, CHENGFEI GUO, TENGFEI WU, AND XIAOPENG SHAO*

*School of Physics and Optoelectronic Engineering, Xidian University, Xi'an 710071, China*
*xpshao@xidian.edu.cn*

**Abstract:** Light passing through scattering media will be strongly scattered and diffused into complex speckle pattern, which contains almost all the spatial information and spectral information of the objects. Although various methods have been proposed to recover the spatial information of the hidden objects, it is still a challenge to simultaneously obtain their spectral information. Here, we present an effective approach to realize spectral imaging through scattering media by combining the spectra retrieval and the speckle-correlation. Compared to the traditional imaging spectrometer, our approach is more flexible in the choice of core element. In this paper, we have demonstrated employing the frosted glass as the core element to achieve spectral imaging. Obtaining the spectral information and spatial information are demonstrated via numerical simulations. Experiment results further demonstrate the performance of our scheme in spectral imaging through scattering media. The spectral imaging based on scattering media is well suited for new type spectral imaging applications.

## 1. Introduction

Spectral imaging has been developed for many years [1] and it plays an important role in varies applications ranging from astronomical imaging to earth observation, and to biomedical imaging [2]. Light passing through the turbid media such as the biological tissues or the ground glass will be strongly scattered and diffused into complex speckle pattern [3,4], which makes it hard to obtain object information. However, the speckle pattern contains almost all the spatial information and spectral information of the objects. Meanwhile, the scattering media can be served as a 2D spectral dispersing element. The spectral information of object becomes disordered. In recent years, several methods have been proposed to recover the spatial information and some impressive results have been achieved. The wavefront modulation technique [5–7] can achieve focusing or imaging through the turbid media by using a spatial light modulator to optimize the wavefront of incident light or measuring the transmission matrix [8,9] of the imaging system. The speckle-correlation based method [10,11] can also realize imaging through thin scattering layers [12–15] by using the "optical memory effect" (OME) [10,12,16]. In some cases, even a single camera image is sufficient to recover the hidden objects. Even though recovering the spatial information of hidden objects from the speckle pattern is realized, retrieving their spectra is still a challenge. In the spectral domain, the detailed distribution of intensity in a speckle pattern after the light passing through the scattering media is frequency-dependent [18,19]. In order to retrieve the spectrum of input light, the spectrum retrieval technique can be applied by storing the different spectral fingerprints in a transmission matrix [18]. A silicon-based spectrometer can realize the spectral information reconstruction by measuring the transmission matrix, but it needs to couple the light into the multimode fiber whose resolution is affected by its length [19]. In other cases, some speckle-based spectrometers have been realized by using the turbid medium or disordered photonic crystal. A recent example showed that the scattering media can be used to achieve spectral imaging [20]. However, this technique suffers from some shortcomings of a semiconductor nanowire mat: (i) It requires a custom scattering media; and (ii) the spatial resolution is affected by cross-talk between adjacent spatial coordinates.

In this paper, we demonstrate an approach to achieve spectral imaging through the scattering media without a custom scattering media and design a dual-arm framework to carry out our method. (In our work, the object, which has a spatially invariant spectrum, is employed.) Inspired by spectra retrieval techniques used in all-fiber spectrometer [18] and the speckle-correlation imaging method [11], we employ the scattering media to simultaneously achieve spectra information and spatial information. For our system, one arm is used to retrieve spectrum by spectra retrieval technique and another arm is used to reconstruct object via speckle-correlation imaging method. A single-mode fiber (SMF) is taken into the spectral arm and it ensures that the system needs to be calibrated only once. Some simulations are carried out to verify spectra retrieval and the speckle-correlation imaging. The results show that spectral imaging through the scattering media is possible by employing the speckle-correlation and the spectral retrieval. At last, experiments are carried out to stress the effectiveness of spectral imaging through the scattering media. Theoretically, the spectral range of our scheme covers from ultraviolet to near-infrared [21,22]. The most important advantage of this approach is that it can perform very well in spectral imaging through the scattering media. In comparison with other traditional imaging spectrometers, our approach is more flexibility in the choice of core element. For instance, spectral imaging based on this approach has been demonstrated using a frosted glass in this paper. This work will be beneficial to the fields of spectral imaging in biomedical and bio-photonics imaging.

## 2. Principal and simulation

### 2.1 Principal

Figure 1 shows the schematic of spectral imaging through scattering media. The input light is collimated by a collimator and fills the entire object. After transmitting through the object plane, the input light is split into two arms (spectra arm and imaging arm) by a beam splitter (BS). In the spectra arm, the light is coupled to the SMF by an objective (OB). The collimated (C) output light of SMF, which transmits through the scattering medium 2 and is captured by the Camera2. In the imaging arm, the light is directly illuminated on the scattering medium 1, and then captured by the Camera1.

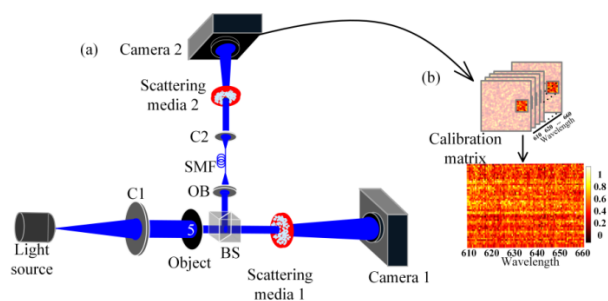

Fig. 1. (a) Schematic of spectral imaging through scattering media: BS, beam splitter, OB, objective, SMF, single-mode fiber, C1 and C2, collimator. (b) The process of calibration matrix.

For spectral retrieval through the scattering media, the system will need to be recalibrated when the object changes. Inspired by optical waveguide mode theory and optical light propagation through the fiber, we employ a SMF in the spectra arm to achieve spectral imaging with calibrating once only. Limited by the optical waveguide mode, the output signal of the single-mode fiber is a certain mode for a certain spectra input light [19]. The calibration matrix is obtained by selecting the same square area of interest (SAI) from a series of speckle patterns such as the ones shown in Fig. 1(b). Calibration light from a Xenon lamp light source is spectrally filtered using a monochromator.

After calibration, the calibration light source can be changed to other light sources. A spectrum retrieval method, to be described in the section 2.1.1, is then applied to retrieve the spectrum of the input. An imaging method, to be described in the section 2.1.2, is then applied to reconstruct the image of the object.

### 2.1.1 Spectra retrieval method

In this section, we describe a spectral retrieval method of reconstructing an arbitrary input spectrum. The speckle patterns generated by single wavelength could theoretically be calculated [23,24], the numerical calculation of the speckle pattern is presented in the Appendix.

In order to resolve an arbitrary input, the input signal's spectrum $s$ is expressed by $\mathbf{I}=\Psi s$ to the acquired speckle pattern $\mathbf{I}$, where $\Psi$ is the calibration matrix. Matrix left-inversion gives a least-square solution $s=\Psi^{-1}\mathbf{I}$, which applies in situations even when $\mathbf{I} \notin span\{\lambda_i\}$ where $\lambda_i$ are the calibrated wavelengths, that is, the spectral retrieval works even for a continuous input spectrum.

Furthermore, the reconstructed spectrum $s_0$ is obtained by solving a general minimization problem [25]:

$$s_0 = \arg\min_{s} \|I - \Psi s\|_2. \tag{1}$$

where $\| \ \|_2$ denotes an $l^2$-norm. Intuitively, $s_0$ represents the reconstructed spectrum. The solution to Eq. (1) is given by:

$$s_0 = VD^{-1}U^T I. \tag{2}$$

where superscript $T$ denotes matrix transpose, and $\Psi = UDV^T$ is the compact singular value decomposition (SVD) of $\Psi$.

By eliminating the non-physical negative solutions and then solving this minimization problem using a minimization optimization algorithm, the spectral retrieval method in Eq. (1) is a minimization problem, we can reconstruct the spectra of a measured signal accurately(CVX [25]is used in our experiments).

### 2.1.2 Speckle-correlation imaging method

In this section, we describe the speckle correlations imaging method. The object is illuminated by a spatially incoherent source. A camera is used behind the scattering medium to capture the transmission light. If the object lies within the range determined by the OME, the system is treated as an imaging system with a spatial shifted-invariant point spread function (PSF) [11,26]. This imaging process can be mathematically represented as the following convolution:

$$I(u) = O(u) * S(u). \tag{3}$$

where the symbol $*$ is a convolution operator, $I(u)$ denotes the detected intensity image, $O(u)$ denotes the object, and $S(u)$ stands for the PSF.

Taking the autocorrelation to both sides of Eq. (3) and using the convolution theorem [11], we obtain:

$$I(u) \bullet I(u) = [O(u) * S(u)] \bullet [O(u) * S(u)] \approx [O(u) \bullet O(u)] + C. \quad (4)$$

where $\bullet$ denotes the autocorrelation operator.

The scattered light autocorrelation, $A(x,y)$, is calculated by an inverse two-dimensional Fourier transform of the camera image's energy spectrum:

$$A(x,y) = I(x,y) \bullet I(x,y) = F^{-1}\left\{\left|F\left\{I(x,y)\right\}\right|^2\right\}. \quad (5)$$

where $I(x,y)$ is the camera image, and $F^{-1}$ and $F$ denote the 2D inverse Fourier transform and the 2D Fourier transform, respectively.

The Fourier phase of the object can be recovered from the Fourier amplitude via a phase retrieval algorithm in the Fourier domain [17].

## 2.2 Simulation

A scattering media model is often necessary in order to simulate the speckle patterns formed through the scattering media. However, according to the principle of speckle phenomena in optics [3], we can simplify the scattering media model as a transmission matrix (phase mask). The angular spectrum theory [24] and the transmittance function [24] are used to simulate the propagation between the different optical components in our simulation. The Primary simulation parameters are listed in Table 1.

**Table 1. Primary Simulation Parameters**

| Parameters | Values | Parameters | Values |
|---|---|---|---|
| Range of wavelength | | | 600~650 nm |
| Spectral sampling interval | | | 0.5nm |
| Spatial sampling interval | | | 14.64 μm |
| Dimension of scattering medium | | | 600×600 |
| Distance between object and scattering medium1 | | | 20 mm |
| Distance between C2 and scattering medium2 | | | 20 mm |

The results of the simulation are shown in Fig. 2 (In the simulation, we assume that the interval between spectral channels is equal to its uncorrelation bandwidth [19]). The pictures in the first row and in the second row display the speckle patterns captured by the camera1 and camera2, respectively. The third row of Fig. 2 show the corresponding results of the spectral retrieval. The fourth row of Fig. 2 show the corresponding results of imaging.

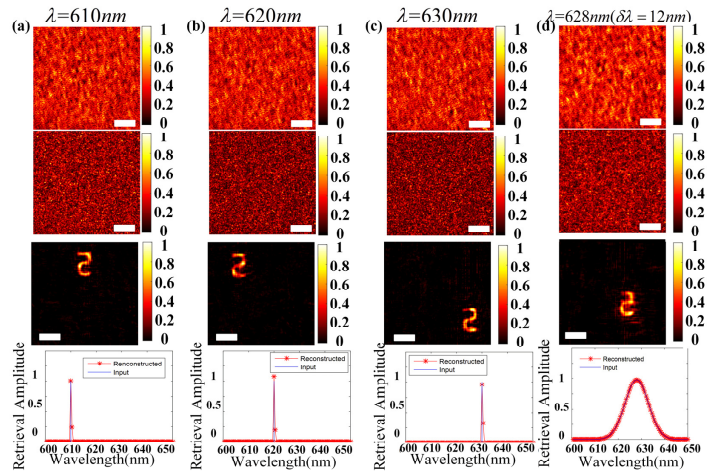

Fig. 2. Numerical simulation results of spectral imaging through the scattering media. (a) First row: speckle patterns captured by camera1 in the condition of the 610nm spectrum illumination. Second row: speckle patterns captured by camera2 in the condition of the 610nm spectrum illumination. Third row: object reconstruction from the first row of (a). Forth row: spectral retrieval from the second row of (a). (b)–(d) Similar to (a) but for different spectral source illuminations. Scale bars: 120 pixels in the first row and second row of (a), (b), (c) and (d), and 12 pixels in the third row of (a), (b), (c) and (d).

As we can see in the third row of Fig. 2(d), the central wavelength of retrieved spectra presented is 628nm, which is perfectly accordant with the central wavelength of input light, of which the central wavelength is 628nm and the FWHW is 12nm. The results show the feasibility of our system.

The spectral retrieval method is sensitive to the noise of optical system [18–20]. In order to estimate the effect of the noise on the retrieved spectra, the Gaussian noise, with an SNR ranging from 50dB to 1 dB (we use the SNR to describe the noise level), is added on the speckle pattern. The Fig. 3 shows the correlation results between the retrieved spectra under different noise levels and the retrieved spectra without noise. As demonstrated in Fig. 3, with the noise increasing from 50 dB to 35 dB, the correlations for both the presented method and Tikhonov regularization (TR)method [19] are near to 1 when SNR>35 dB, and rapidly drop when SNR≤35 dB. It can be inferred that the anti-noise performance of the proposed method is superior to TR method in our simulation, which indicates its great latent capacity for retrieving the spectra through the turbid media in a noise environment. (TR requires careful adaptation of the noise level. While both methods are able to accurately reconstruct wavelengths, CVX is able to minimize reconstruction noise while TR produces many small values fluctuating around zero.)

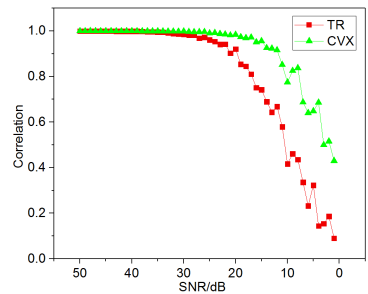

Fig. 3. Performance of spectral retrieval with increasing noise.

## 3. Experiment and analysis

### 3.1 Experimental

The experimental optical setup is shown in Fig. 4(a). We use Thorlabs DG10-220 diffuser with 2mm thickness, 220 grit, as the scattering medium. The objects are digit characters, which are filtered from a resolution test target (1951USAF, Edmund Company).

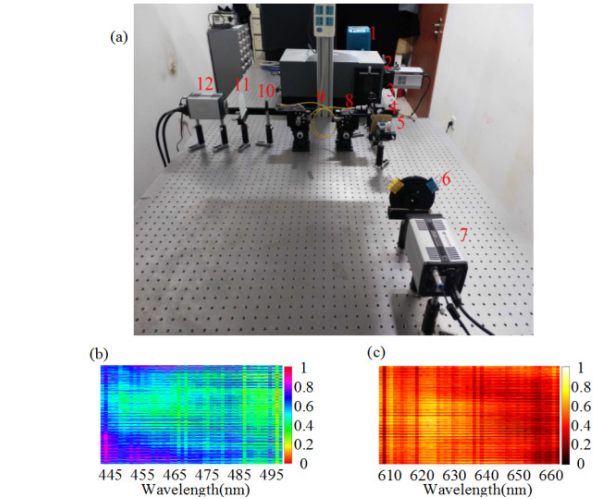

Fig. 4. Experimental setup and calibration matrices of spectral imaging through scattering media. (a) Experimental optical setup: 1, Xenon lamp light source; 2, monochromator; 3, collimator; 4, BS; 5, object; 6, scattering media1; 7, camera1; 8, OB; 9, SMF; 10, C2; 11, scattering media2; 12, camera1. (b,c) Calibration matrix showing the intensity distributions of different wavelengths through the experiment system from 445 to 495 nm and from 610 to 660 nm.

The spectra arm needs to be pre-calibrated. Calibration light from a Xenon lamp light source is spectrally filtered with a monochromator, resulting in a tunable source with spectral resolution of 1 nm (full-width half-maximum) FWHM. The monochromator (Andor Spectrograph, Shamrock 500i), consisting of a 1200 lines/mm grating and a Tungsten-Halogen lamp source (Zolix, GLORIA-X500A), produces a 1.0 nm FWHM probe signal to extend our previous analysis to locate in the range of $445nm < \lambda < 495nm$ and $610nm < \lambda < 660nm$. The CMOS camera (AndorZyla5.5) with 12-bit, 4.2M pixels and 6.5 μm pixel size is used to capture the speckle pattern.

The calibration method described in Sec. 2.1 [see the process in Fig. 1 (b)] is employed to calibrate the spectral arm. The calibrated matrices are shown in Figs. 4(b) and 4(c). The wavelengths difference between each speckle pattern is 1 nm.

To investigate the ability of our system to achieve the spatial and spectral information, different objects and different optical sources are employed in our experiment. Figures 5(a) and 5(c) show the experimental results of spectral imaging by narrow bandwidth light source.

To test the validity of broadband continuous spectra, the broadband continuous spectral sources are used to replace the narrowband source. The corresponding experimental results are shown in Figs. 5(b) and 5(d).

As demonstrated in the first row of Fig. 5(a), the reconstructed object is presented in the pictures in the first row and the original object is in the bottom left. The reconstructed object is identical in the spatial information with the original object. As show in the second row of Fig. 5(a), the central wavelength of retrieved spectra is 470nm and the central wavelength of input light is 470nm. Figure 5(b), Fig. 5(c) and Fig. 5(d) are similar to Fig. 5(a) expect for different source illuminations or different objects. The results indicate that our structure can achieve spectral imaging through the scattering media.

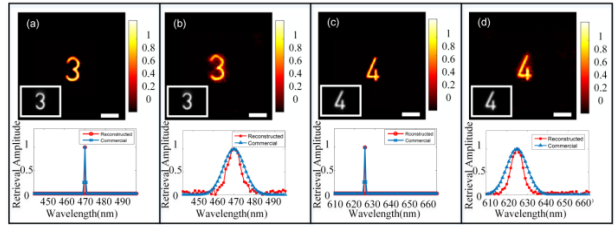

Fig. 5. Reconstructing spatial and spectral information. (a) First row: Reconstructed spatial intensities from the speckle pattern captured by camera1. Second row: Spectra recovered from the speckle pattern captured by camera2. (b)–(d) Similar to (a) but for different source illuminations or different objects, respectively. Scale bars: 52 um in the first row of (a), (b), (c) and (d).

### 3.2 Analysis of spectral retrieval

In order to evaluate the effectiveness of our system for continuous spectra light source, we employ the LED light source as the probe signal. The solid line represents the spectral signal measured by a commercial grating spectrometer and the dotted line represents the retrieved spectra, [see Figs. 5(b)–5(d)]. As demonstrated in Fig. 5(b) [Fig. 5(d)], the central wavelength and FWHM of input light is 470nm and 16nm (625nm and 16nm), respectively. The central wavelength and FWHM of retrieved spectra is 470nm and 14nm (625nm and 14nm) in Fig. 4(b) [Fig. 4(d)].

It can be deduced from Fig. 5 that our experiment realizes the spectral retrieval for narrow bandwidth and continuous spectra light sources, which implies the capacity of our method and experiment setup for retrieving the spectra through the turbid media in a practical application.

On the basis of the experimental scheme discussed in the previous section, we can reconstruct the spectrum of multiple probe beams. Figures 6(a) and 6(b) show the reconstruction results of a different range of wavelengths for a narrowband spectrum signal. Figures 6(c) and 6(d) show the experiment results of spectral retrieval for the narrow bandwidth light sources. As a contrast, pictures in the second row of Figs. 5(b) and 5(d) show the experiment results of spectral retrieval for the continuous spectra light sources.

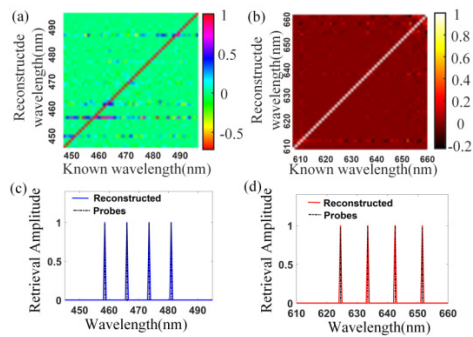

Fig. 6. Experimental results of spectral retrieval. (a,b) Spectral retrieval of a different range of wavelengths for narrow spectrum; (c,d) Comparing the results of spectral retrieval with the probes signals. (Narrow-band light sources are used to individually illuminate the sample.)

In Fig. 6(c), the black dashed lines mark the central wavelengths of the input light sources and the blue solid lines mark the retrieved spectra. The Fig. 6(d) is similar to Fig. 6(c), except for the red solid lines marking the retrieved spectra. So, the central wavelengths of spectra are retrieved accordance with the central wavelengths of input light source.

### 3.3 Analysis of speckle-correlation imaging

Figure 7 shows the experiment results of speckle-correlation imaging method. The original objects are shown in the first row of Figs. 7(a)–7(e). The reconstruction object results at 1nm FWFM and 16nm FWFM are shown respectively in second row and third row of Fig. 7. Compared with the results in the second row, the results in the third row show some deviations when the FWHM increasing from 1nm to 16nm.

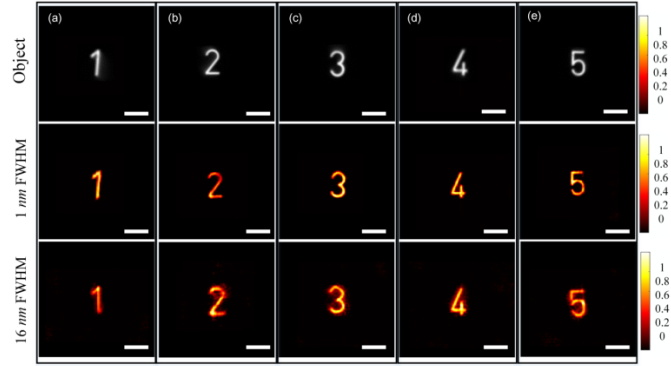

Fig. 7. Experimental results of speckle-correlation imaging. (a) First row: original object. Second row: reconstruction image from speckle patterns in the condition of 1 nm FWHM. Third row: reconstruction image from speckle patterns in the condition of 16 nm FWHM. (b)–(e) Similar to (a) but for different objects. Scale bars: 130 um in the first row of (a), (b), (c) and (d), and 195 um in the second row and third row of (a), (b), (c) and (d).

One main reason for the deviations is the low speckle correlation caused by broadband spectrum of light source. Theoretically, the Eq. (4) in section 2.2 is correct in the condition of monochromatic light source. With increasing the bandwidth of light source, the Eq. (4) can be reformulated as:

$$I(u) \bullet I(u) = [O(u) * S(u)] \bullet [O(u) * S(u)]$$
$$= [O(u) \bullet O(u)] * \left[ \sum_{i=1}^{M} S_{\lambda_i}(u) \bullet S_{\lambda_i}(u) + \sum_{i=1}^{M} \sum_{j \neq i}^{M} S_{\lambda_i}(u) \bullet S_{\lambda_j}(u) \right]. \quad (6)$$

where $S_{\lambda_i}(u)$ denote the PSF of $\lambda_i$, $\lambda$ represent the wavelength. Compared with the Eq. (4), the Eq. (6) has an extra item $\sum_{i=1}^{M} \sum_{j \neq i}^{M} S_{\lambda_i}(u) \bullet S_{\lambda_j}(u)$. In the narrow bandwidth spectra of light source, the value of $\sum_{i=1}^{M} \sum_{j \neq i}^{M} S_{\lambda_i}(u) \bullet S_{\lambda_j}(u)$ is much less than that of $\sum_{i=1}^{M} S_{\lambda_i}(u) \bullet S_{\lambda_i}(u)$, so the effect of $\sum_{i=1}^{M} \sum_{j \neq i}^{M} S_{\lambda_i}(u) \bullet S_{\lambda_j}(u)$ can be ignored. In contrast, for broadband continuous spectra input light, the item $\sum_{i=1}^{M} \sum_{j \neq i}^{M} S_{\lambda_i}(u) \bullet S_{\lambda_j}(u)$ plays an important role in the object reconstruction process and cannot be ignored. Based the analysis above, the quality of object reconstruction will become poor with the increasing of bandwidth. As shown in Fig. 6, the results become fuzzy with the increasing of bandwidth.

In order to evaluate the imaging quality of objects in different FWHM illumination light, the correlation index is imported. It represents the correlation between the reconstruction object and original object. From the correlation indices shown in Table 2, the correlation index show decreasing with the increasing of bandwidth.

Table 2. Quantitative Comparison for Experiment Results (Correlation)

| Object | 1 | 2 | 3 | 4 | 5 |
|---|---|---|---|---|---|
| 1 FWHM | 0.9817 | 0.9856 | 0.9848 | 0.9824 | 0.9882 |
| 16 FWHM | 0.9420 | 0.9569 | 0.9565 | 0.9545 | 0.9651 |

### *3.4 Discussion*

Several important points related to our work should be discussed in detail. Firstly, in our experiments, we employ a spectrum to represent the spectral information of object. Due to the random dispersion of scattering media, the spectral information of object becomes disordered in the speckle pattern. Limited by the speckle correlation imaging method, the positional information of object is uncertain. Thus, the reconstructed spectrum stands for the spectral information of object and could be used to represent the spectral information of object. Secondly, Spectral resolution is an important parameter in our method. Although the study has successfully demonstrated that we can get the spatial and spectral information of object behind the scattering media, it has certain limitations in terms of spectral resolution. At the present, by the limitation of instrument, we have not the capability of testing the spectral resolution. Further investigation and experimentation related to the spectral resolution of our method is strongly recommended. Finally, we only realized the spectral imaging of transmission object in our experiments. Our method can also be extended to reflective object. The method paves the way for acquiring the image and spectral information of objects in real time in the fields of biomedical and bio-photonics imaging.

### 4. Conclusion

In this paper, we propose an effective approach to realize spectral imaging through scattering media and put experiment on it. To demonstrate the feasibility of the proposed method, spectral imaging through the scattering media is simulated. And the results show that the spatial and spectral information can be achieved simultaneously. Experiments using the frosted glass, further demonstrate the ability of our method for both narrow bandwidth light sources and continuous spectra light sources. However, because of the presence of background light and noise, the experiment results are not as accurate as those in the simulations, and further efforts are needed to improve performance of spectrum retrieval and object reconstruction in practice. Spectral imaging system based on scattering media can have much lighter weight, smaller size, and lower cost than traditional grating or prism based imaging spectrometers. This work will be beneficial to developing the new type spectral imaging system.

### Appendix

Here we present the analytical calculation of the speckle pattern. The setup is shown in Fig. 8, where the distances from the object to turbid medium and lens are $d$ and $S_o$, respectively. The camera is located in a distance $S_i$ from the lens.

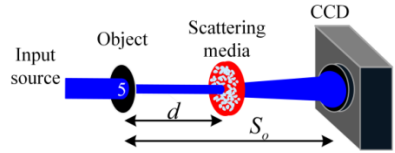

Fig. 8. Schematic diagram of scattering system.

The performances of our imaging system can be analyzed using Huygens-Fresnel field propagation. The field at the image plane in the system can be described as [23–26]:

$$E(\rho_I, \lambda) = A \iint E(\rho_O, \lambda) e^{\frac{ik}{2d}(\rho_T - \rho_O)^2} Pup.(\rho_T, \lambda) T(\rho_T, \lambda) e^{\frac{ik}{2(S_O - d)}(\rho_I - \rho_T)^2} d\rho_O d\rho_T \quad (7)$$

Here, $k = 2\pi/\lambda$. $\rho_O, \rho_T, \rho_I$ are the coordinates in the object plane, scattering plane medium plane and image camera) respectively. We denote the generic complex function of the turbid medium operator by $T(\rho_T, \lambda) = t(\rho_T, \lambda) \cdot e^{ikh(\rho_T)}$, and a proportionality constant by $A$. $Pup.$ is the explicit pupil function as limiting aperture of the system, $h$ stands for the surface height of scattering media at the micro-scale.

Considering the effects of wavelength, the intensity in the camera plane can be expressed as

$$I(\rho_I, \lambda) = A^2 \left| \iint E(\rho_O, \lambda) e^{\frac{ik}{2d}(\rho_T - \rho_O)^2} Pup.(\rho_T, \lambda) t(\rho_T, \lambda) e^{ikh(\rho_T)} e^{\frac{ik}{2(S_O - d)}(\rho_I - \rho_T)^2} d\rho_O d\rho_T \right|^2. \quad (8)$$

Equation (8) simplifies to:

$$I(\rho_I, \lambda) = A^2 \left| \iint E(\rho_O, \lambda) \beta(\rho_I, \rho_T, \rho_O, \lambda) d\rho_O d\rho_T \right|^2. \quad (9)$$

where $\beta(\rho_T, \rho_O, \lambda) = Pup.(\rho_T, \lambda) t(\rho_T, \lambda) e^{\frac{ik}{2(S_O - d)}(\rho_I - \rho_T)^2} e^{\frac{ik}{2d}(\rho_T - \rho_O)^2} e^{ikh(\rho_T)}$.

To broadband continuous spectrum, the intensity in the camera plane can be expressed as

$$I(\rho_I) = \int I(\rho_I, \lambda) d\lambda. \quad (10)$$

In our simulation, we regard $t(\rho_T, \lambda)$ as a constant.

## Funding

## Acknowledgment

The authors thank Yuxiang Wu and Bingxin Tian for insightful discussions and support with writing this paper.